\documentclass[usenatbib,preprint,twocolumn]{mn2e}
\usepackage[utf8]{inputenc}
\usepackage{graphicx}
\usepackage{graphics}
\usepackage{multicol}
\usepackage{color}
\usepackage{tabularx}
\usepackage{longtable}
\usepackage{times}
\newcommand{\software}[1]{\texttt{\small \MakeUppercase{#1}}}
\usepackage{fancyhdr}
\fancypagestyle{firststyle}
{
   \fancyhf{}
   \fancyhead[L]{MNRAS \textbf{458}, 1302-1310 (2016)}
   \fancyhead[R]{doi:10.1093/mnras/stw290}
   \fancyfoot[R]{$\textcopyright$ 2016 The authors \\ Published by Oxford University Press on behalf \\ of the Royal Astronomical Society}
   
}

\title{SALT observation of X-ray pulse reprocessing in 4U 1626-67$^{\ast}$}
\author
{Gayathri Raman$^{1\dagger}$, Biswajit Paul$^{1}$, Dipankar Bhattacharya$^{2}$ and Vijay Mohan$^{2}$ \\ 
$^1$Department of Astronomy and Astrophysics, Raman Research Institute, Bangalore 560080, India\\
$^2$Inter-University Center for Astronomy and Astrophysics, Pune 411007, India\\
}

\begin{document}
\date{Accepted 2016 February 3. Received 2016 February 3; in original form 2015 June 12}
\maketitle
\thispagestyle{firststyle}

\begin{abstract}
We investigate optical reprocessing of X-rays in the LMXB pulsar 4U 1626-67 in its current
spin-up phase using observations with Southern African Large Telescope (SALT), near-simultaneous observations 
with Swift-XRT and non-simultaneous \textit{RXTE}-PCA observations and present the results of timing analysis. 
Using SALT observations carried out on 2014, March 5 and 6, we detect some interesting reprocessing signatures.
We detect a weak optical Quasi Periodic Oscillation (QPO) in the power density spectrum on March 5 at 48 mHz 
with a fractional rms of 3.3$\%$  in spite of the fact that source shows no corresponding X-ray QPO in the spin-up phase. 
In the light curve obtained on March 5, we detect a coherent pulsation at the spin period of $\sim$7.677 s.
A previously known, slightly down-shifted side-band is also detected at 129.92 mHz. The frequency spacing between main 
pulse and this side-band is different from earlier observations, though the statistical significance of the difference is limited.
The light curve of March 6 displays short time-scale variability in the form of flares on time-scales of a few minutes.
Folded pulse profiles resulting from data of this night show an interesting trend of pulse peak drifting.
This drift could be due to i) rapid changes in the reprocessing agent, like orbital motion of an 
accretion disc warp around the neutron star or ii) intrinsic pulse phase changes in X-rays. We also examine some X-ray
light curves obtained with RXTE-PCA during 2008-2010 for pulse shape changes in short time-scales during X-ray flares.
\end{abstract}

\footnotetext[1]{based on observations made with the Southern African Large Telescope (SALT)}
\footnotetext[2]{graman@rri.res.in}

\section{Introduction}
 Reprocessing of X-rays into the visible and UV bands is a useful diagnostic tool to characterize and map the surrounding medium 
 responsible for reprocessing and also to investigate the physics of reprocessing. 
 In Low Mass X-ray Binaries (LMXBs), the accretion disc dominates the optical emission, whereas in High Mass
 X-ray Binaries (HMXBs), the massive companion star overwhelms the optical emission. Thus, LMXBs present the right setting to
 explore the matter distribution in regions around the neutron star. \\
 LMXBs show many interesting X-ray timing variabilities. These X-ray variabilities sometimes get reproduced in the Optical/UV band 
 through reprocessing. There are several possible reprocessing sites, namely different parts of the accretion disc, 
 the surface of the companion star facing the neutron star, or sometimes both. The observed reprocessed emission could 
 be modulated at the orbital period while different portions of the X-ray heated region come in and out of view. 
 These optical signatures of emission processes could get delayed, smeared or may also possess a phase shift with respect to
 the X-ray variability. A number of sources show correlated thermonuclear bursts in X-rays and Optical
 (EXO 0748-676: \citealt{hynes,bpaul}). Some LMXBs show Quasi Periodic Oscillations (QPOs) in their X-ray Power Density Spectra (PDS)
 that are also seen in the optical band (4U 1626-67: \citealt{deepto1998}; GX 339-4 \citealt{mci}). Some X-ray pulsar sources show
 coherent optical pulsations (GX 1+4: \citealt{jablonski}; 4U 1626-67 \citealt{middleditch}) indicating that a large fraction 
 of their optical emission is due to reprocessing. Then there are sources that show correlated flaring behavior 
 (4U 1626-67: \citealt{jar,mcclintock1980}). Extensive study on broad and narrow, fast and slow X-ray and optical 
 flares and their mutual time lags in some black hole sources has been carried out by \citet{gandhi}.
 Co-variability studies between the X-rays and optical are thus crucial in understanding different scenarios of X-ray reprocessing. \\
 4U 1626-67 has a magnetic field strength of about 3x10$^{12}$ Gauss, estimated using the cyclotron resonance scattering 
 feature detected in the X-ray spectrum (\citealt{orlandini,cameroarranz}). The pulsar has a spin period of about 7.7 s 
 with no evidence of pulse arrival time delay due to orbital motion and at the 42 min orbital period measured from an optical 
 sideband by \citet{middleditch}, the upper limit of the projected semi-major axis is 10 lt-ms \citep{rappaport,levine,jain2008}. 
 Based on these, and faintness of the optical counterpart the neutron star was understood to have an extremely low mass and 
 degenerate companion star \citep{levine,deepto1998}. It has one of the lowest mass functions of
 $<$1.3 x 10$^{-6}$ M$_{\odot}$ \citep{levine} as measured for any X-ray binary. A very low inclination or a nearly face 
 on geometry (\textit{i}$<$10$^{\circ}$) allows a higher mass companion although the a priori probability of finding such a system is 
 only 0.015. The lower limit for the mass transfer rate has been calculated by \citet{deepto1998} as 2x10$^{-10}$ M$_{\odot}$ per year,
 using the maximum possible torque that can act on the neutron star during spin-up, which is when the magnetospheric radius comes 
 very close to the co-rotation radius. \citet{deepto1998} showed that a 0.08 M$_{\odot}$ hydrogen-depleted companion, 
 with an \textit{i}$<$8$^{\circ}$ (apriori probability of $\sim$ 1$\%$) would satisfy the minimum mass transfer rate criterion 
 and would also be at a reasonable distance of 3 kpc. The system shows no type 1 X-ray thermonuclear bursts, but does show
 frequent X-ray flares. \\
 Most accretion powered X-ray pulsars show torque reversals in time-scales of weeks to months and years that is often related to 
 the X-ray luminosity \citep{bildsten}. Not only does 4U 1626-67 not show the standard X-ray luminosity- accretion torque 
 relation \citep{beri}, but it has undergone only two such torque reversals in the last 40 years \citep{deepto1997,jain,cameroarranz}.
 The source shows different X-ray features during the spin-up and spin-down eras. For example, the X-ray pulse profiles are 
 energy dependent and show a double horned shape during the two spin-up eras, which gets turned into a dip during the
 spin-down era \citep{beri}. The strong 48 mHz QPO found consistently during the spin-down phase \citep{kaur}, 
 is absent in the power spectrum of the current spin-up phase \citep{jain}, although a weak X-ray QPO was reported
 by \citet{shinoda} during the first spin-up epoch. The X-ray spectra are similar during the spin-up eras
 as against the spin-down era \citep{cameroarranz}. All of these indicate that there are changes in the accretion mode from 
 the inner disc to the neutron star that are associated with the torque reversals \citep{beri}.\\
 The very faint blue optical counterpart KZ Ter was discovered by \citet{mcclintock1977} and was found to have an UV excess
 in its spectrum. The magnitudes obtained with the identified source were not consistent with a main sequence or supergiant
 companion, but were more likely supportive of a system containing a late type dwarf or a degenerate dwarf where most
 of the optical light came from reprocessed X-ray flux by an accretion disc and/or the inner face of the companion
 star \citep{jar}. After that, optical pulsations were detected by \citet{IlovaskyMotchChev} at the same X-ray pulse
 period of 7.67 s with a pulsed amplitude of 4$\%$ and pulsed energy fraction of 2$\%$. \citet{crampton} established that 
 if the system was an ultra compact binary, then the X-ray heating of the companion must be negligible and that the UV excess
 seen in the source must be associated with an accretion disc \citep{chester}. \citet{mcclintock1980} discovered intense 
 correlated X-ray and optical flares and also suggested that a large fraction ($>$8$\%$) of the pulsed emission is produced 
 within 0.5 lt-s from the Neutron star. This again confirmed the fact that the accretion disc could be the source of 
 the reprocessed optical pulsations. They found a 6.2$\%$ pulsed amplitude, and that the optical pulsations were mainly
 driven by the 1-3 keV soft X-ray pulses.\\
 In the first spin-up era, along with the main pulsation, a sideband of a slightly lower frequency was observed and
 explained as reprocessed pulsation from the donor's surface \citep{middleditch} that was beating with the 
 main pulsation at the orbital period. An orbital period of 42 min and a pulsed fraction of 2.4$\%$ was obtained.
 The sideband and orbital period was confirmed by \citet{deepto1998}. The fact that the side band occurred at a lower
 frequency indicated that the pulsar spin was in the same direction as the orbital motion. After the first torque reversal,
 in the spin-down era, \citet{deepto1998} reported a 48 mHz optical QPO with a fractional rms amplitude of 6-8$\%$ in 
 the power density spectrum and a pulsed fraction of 6$\%$. Another 1 mHz QPO was detected by \citet{deepto2001} without 
 a simultaneous X-ray QPO at that frequency. This was associated with a possible warp in the accretion disc. \\
 Detailed observations in the last four decades have indicated change in accretion mode associated with change in accretion
 torque \citep{jain,beri}. In this paper we investigate the optical temporal characteristics of 4U 1626-67 after the 
 second torque reversal, when it is in a spin-up phase to re-examine various features like the relative pulse
 amplitudes in X-rays and optical, the strength of the previously reported orbital sidebands and the pulse profile evolution.
 New observations were carried out with the Southern African Large Telescope (SALT) telescope, \textit{Swift}-XRT and we have also done analysis of archival X-ray 
 data to examine the pulse characteristics during X-ray flares.
  
\section{Observations and Data Analysis}

\subsection{SALT}
\subsubsection{Observations}
4U 1626-67 was observed using the SALT for two consecutive nights on 2014 March 5 and 6. 
The SALT telescope is the largest ground-based optical telescope operating in the Southern hemisphere with a 11 m 
spherical primary. It is fixed at a constant elevation angle at 37$^{\circ}$ from the vertical \citep{buckley,donoghue}.
So it can observe only a given annulus of the sky. This gives an observing time of nearly 2 hours twice as the source drifts 
through the annulus \citep{gunther}.\\
The observations were made using the SALTICAM instrument. Details of the observations are given in Table 1. The SALTICAM 
was operated using the \textit{V} band filter (central wavelength 550 nm). The slot mode gives a fast readout time with practically
no dead time, and so, with a 3x3 pixel binning, a sampling time of nearly 0.2 s was achieved. The clocks used for time sampling
produced a binning of 0.203 or 0.204 s. For a very small fraction of the observation, the binning is uneven, separated by
0.1 or 0.6 s. We note here that the instrument configuration was identical on the two nights and therefore the difference
between the results obtained on the two nights of observation are not due to any instrumental artefact.
The photons are collected by two CCDs separated by 1.5 mm, each with two readout amplifiers. The readout noise is 
less than 3 electrons pixel$^{-1}$. The gain specified was 2.6 electron per ADU.

\begin{table*}
\centering
\begin{center}
 \resizebox{1.4\textwidth}{!}
 {\begin{minipage}{\textwidth}
\caption{Log of all optical and X-ray observations of 4U 1626-67 used for this work.} 
\begin{tabular}{cccc}
\hline
 Observation & Date & Exposure time (ks) & Observation length (ks)  \\
\hline
SALT night 1 & 2014 March 5 & 4.8 & 4.8\\
SALT night 2 & 2014 March 6 & 6.6 & 6.6\\
\textit{Swift}-XRT (00031156002) & 2014 March 5 & 4.6 & 4.6\\
\textit{RXTE}-PCA (95313-01-01-08) & 2010 December 9 & 18.2 & 20.3 \\
\textit{RXTE}-PCA (95338-05-01-00) & 2010 January 14 & 10.9 & 17.1\\
\textit{RXTE}-PCA (95338-05-02-00) & 2010 January 15 & 7.3 & 9.6\\
\hline 
 \end{tabular}
 \end{minipage}}
 \end{center}
 \end{table*}
 
\subsubsection{Data reduction}
The bias subtracted SALT data were first extracted using the package \software{IRAF}. The target star and two other comparison 
stars were chosen from a single amplifier. The image frames that contained these stars were then further reduced. The overscan 
regions in the raw images were trimmed. An aperture photometry was then carried out on these images using the \software{SExtractor} 
software. The magnitude and fluxes of the target star and comparison stars were extracted using an aperture with inner and
outer radii as 16 and 25 pixels, respectively. The point spread function had a full width at half-maximum of 3 arcsec. 
These images were then used to remove all the identified
stars and create a flat image. Around 50 flat images were averaged to produce a master flat, which was then used for 
flat fielding the images. The flat fielded, bias subtracted images were then used for determining the magnitudes and the 
fluxes of the target and comparison stars. In some of the exposures, the algorithm could not identify the star and hence the
light curve had certain individual bins that had very low counts. All such bins ($<$2$\%$ of the total bins) were manually removed.

\subsubsection{Timing analysis}
Figure 1 shows the light curves for both the nights with a bin size of 7 s. The March 5 light curve shows slow variations
around an average rate. The optical intensity varies by about a factor of 2 on the first night and by a factor of 3 on the second night.
The March 6 light curve shows many flares of duration between 3 and 15 minutes recurring
at shorter time-scales during some parts of the light curve. The flaring amplitude is about 1.5 to 2 times above the quiescent levels.
These flares are of shorter duration with shorter recurrence time-scales 
as compared to earlier reports (Joss et al. 1978), which had ~500s  co-incident X-ray and optical flares, recurring every ~1000 s. 
In the spin-down phase of 4U 1626-67, it is known to show no X-ray flares \citep{deepto2001,beri}. \\
The PDS generated from the two nights are shown in Figure 2.
The PDS was normalized such that its integral gives the squared rms fractional variability and it was also white noise subtracted.  
The PDS has a power law shape and shows the pulse peak and its harmonics. 
The March 5 night data, surprisingly, shows a weak broad QPO-like feature at nearly 48 mHz.
This PDS was fit with a power law in the frequency range 2 to 200 mHz, excluding the frequency bins near the pulse frequency 
and the harmonics and an excess in power around 0.048 Hz was fit with a Lorentzian centred at that frequency.
The optical QPO has a fractional amplitude of 3.3$\%$ (Figure 2). We also see a broad feature 
centred at 3.2 mHz in the PDS of the March 6 night data. This could be the signature 
of the flares that are present in the light curve.\\

\begin{figure}
 \centering
 \includegraphics[scale=0.3, angle=-90]{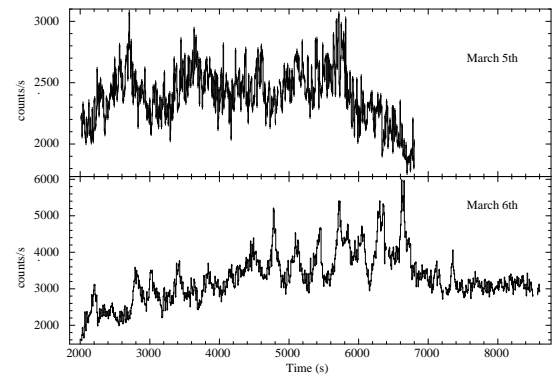}
 \caption{ V-band light curves of 4U 1626-67/KZ TrA with a bin size of 7 s shown for 2014 March 5 (top) and March 6 (bottom). 
 March 6 light curve shows a number of flaring episodes. }
\end{figure}

\begin{figure}
\centering
\begin{minipage}{1.0\linewidth}  
\centering
  \includegraphics[scale=0.4,angle=-90]{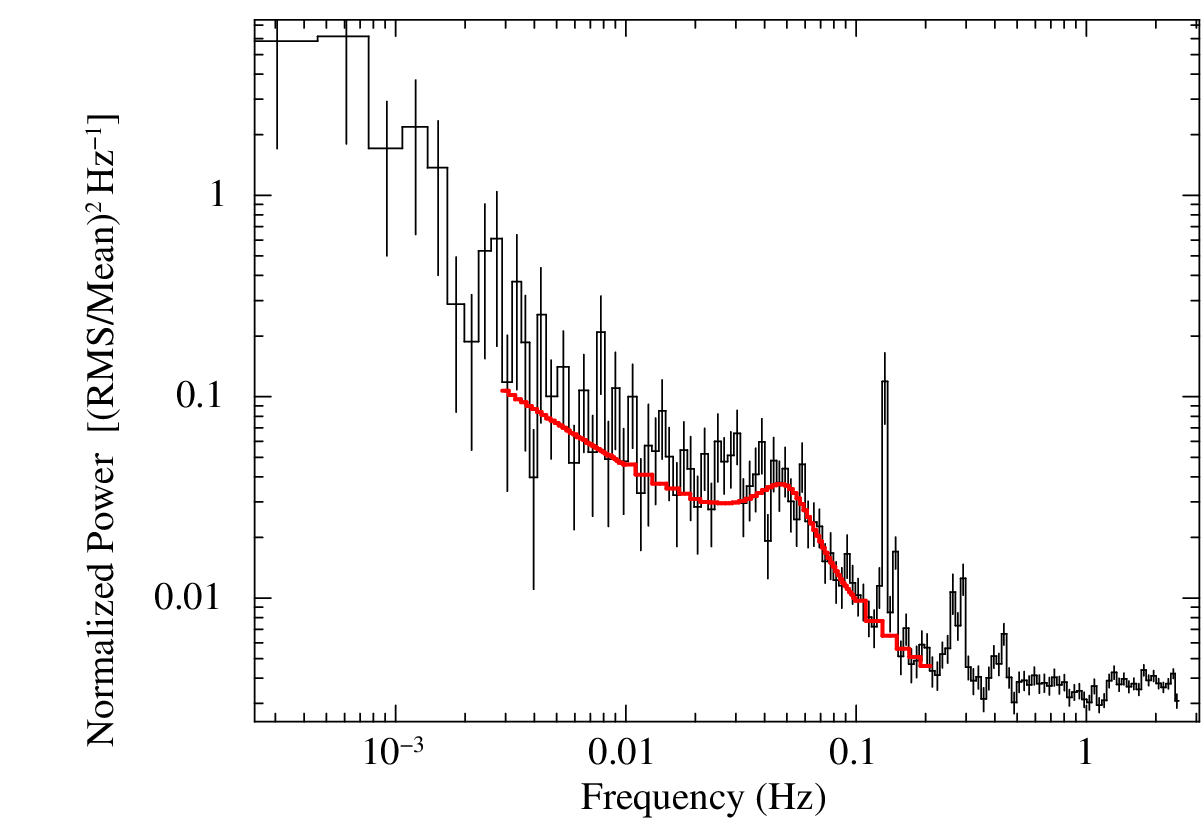}
  \end{minipage}
  \begin{minipage}{1.0\linewidth}  
  \centering
 \includegraphics[scale=0.4,angle=-90]{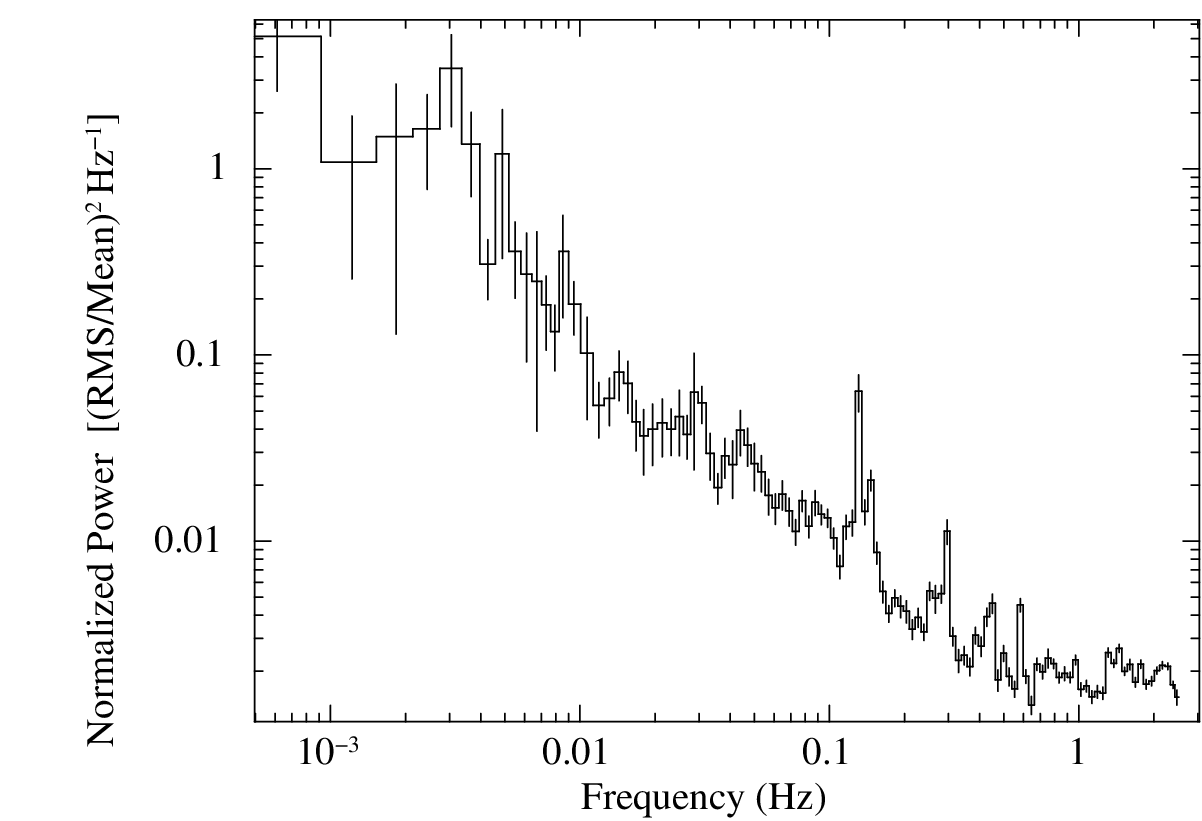}
 \end{minipage}
 \caption{White noise subtracted PDS of March 5 (top) and 6 (bottom) night data normalized relative
 to the mean source power. The broad 48 mHz optical QPO detected in the March 5 SALT PDS is fit with a power law and a Lorentzian
 and shows a fractional rms amplitude of 3.3$\%$. }
 \end{figure}

The optical pulse period was searched for using trial periods around the known X-ray pulse period of 7.6736 s obtained from 
the near simultaneous \textit{Swift}-XRT observations (see the next section).
The \software{FTOOL} \software{efsearch} is a time domain epoch folding procedure that is used to determine periodicity accurately
in a time series once an approximate period value is known. Each folded time series is fit with a constant function and 
the best period is obtained by chi-squared maximization. The SALT time series were folded with 1000 different
periods around a value of 7.67 s with a resolution of  0.1 ms and 32 phase-bins per period.
A sharp peak at 7.677 s (130.259 mHz) and a sideband at 7.697 s was obtained on the March 5.
The left-hand panel of Figure 3 shows the \software{efsearch} results for March 5 light curve.
The sideband frequency, measured from the SALT light curve is 129.921 mHz giving a frequency down-shift of 0.338 mHz and 
the corresponding beat period of $\sim$49.3 min as compared to the earlier measured value of 42.2 min or a difference of
0.395 mHz \citep{deepto1998}. The amplitude of the side-band relative to the main pulsation is $\sim$30$\%$, compared to 20$\%$ measured 
by \citet{middleditch} and 25$\%$ measured by \citet{deepto1998}. \\

\begin{figure*}
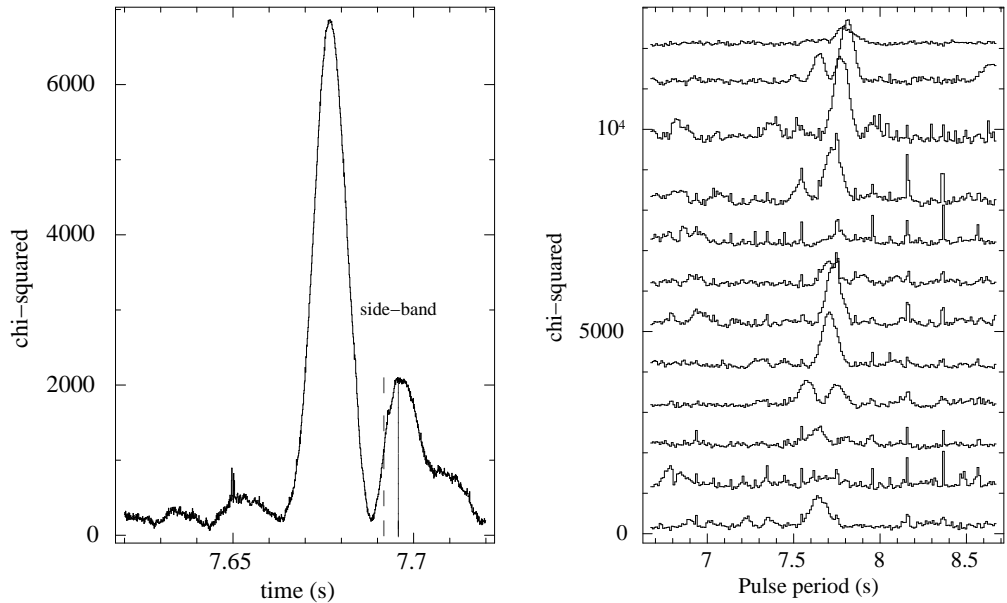

\centering
 \includegraphics[scale=0.45, angle=-90]{efsearch_night1_final.ps}
 \includegraphics[scale=0.45, angle=-90]{efsearch_night2_stacked_final.ps}
\caption{Period search results for March 5 (left) and 6 (right) light curves. A clear side-band peak is seen on March 5 data 
at 7.697 s. The earlier reported side-band is shown by the dashed line. The March 6 light curve has been divided into smaller
segments and epoch folding period searches yield variable pulse periods along the course of the observation. }
\end{figure*}

\begin{figure}
\centering
 \includegraphics[scale=0.3, angle=-90]{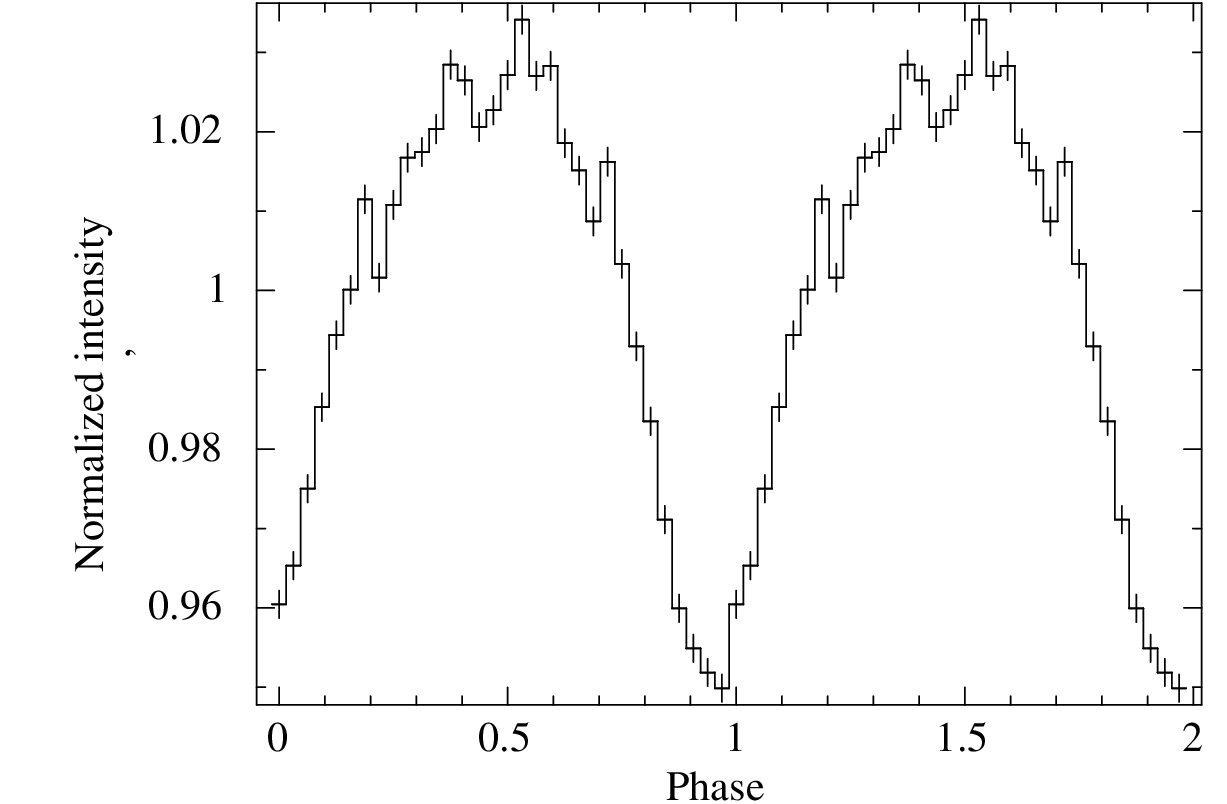}
 \includegraphics[scale=0.3, angle=-90]{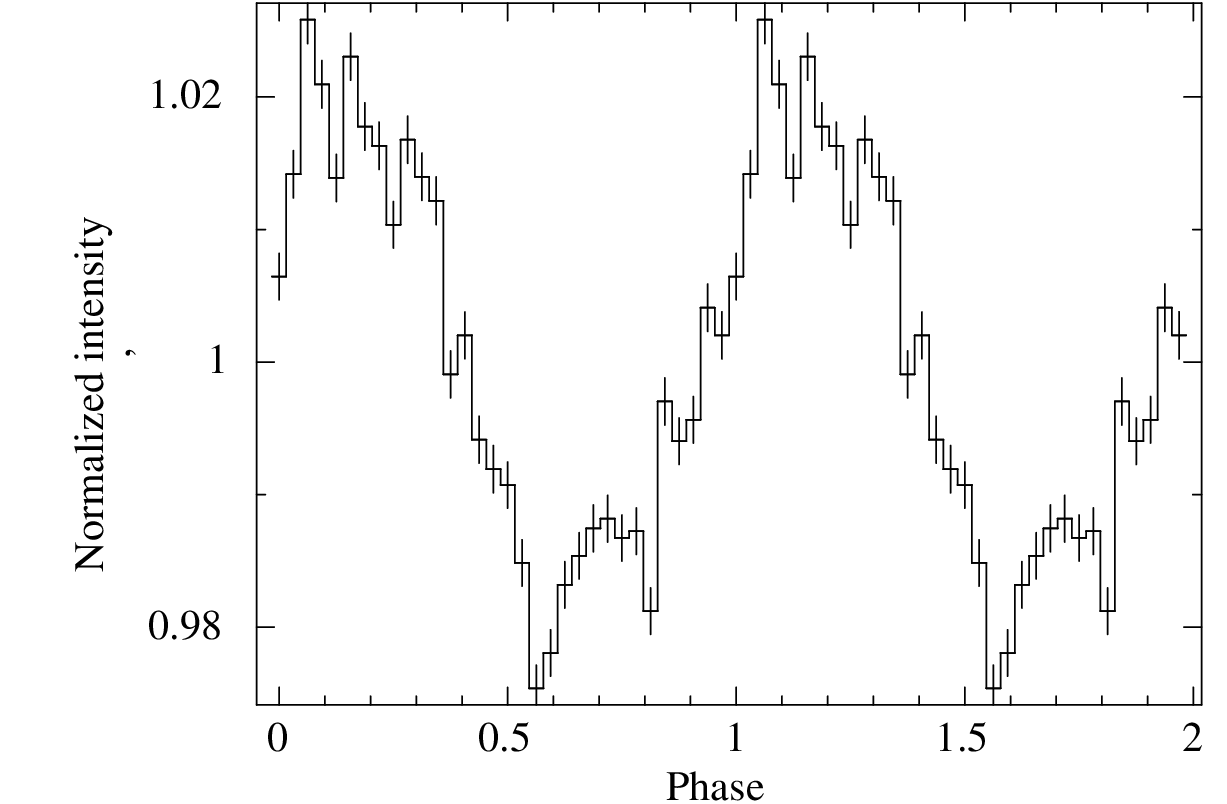}
 \caption{Average pulse profiles for March 5 night folded on the main pulse period 7.677 s (top) and sideband period 7.697 s 
 (bottom).}
\end{figure}

 The average pulse profiles folded at the pulse period and sideband period are shown in Figure 4. The profile shows a 
 single broad peaked feature. The pulse fraction obtained from the average pulse profile on March 5 night was 5$\%$.
 An orbital phase resolved analysis was carried out using the March 5 light curve. The light curve was divided into four 
 orbital phases corresponding to the beat period obtained on night 1 (0.338 mHz or 49.3 min), and separate pulse profiles 
 were created from each segment by folding with the same pulse period. The four resultant profiles are shown in Figure 5. 
 The pulse profile shape clearly varies as a function of orbital phase. \\
 
\begin{figure}
\centering
 \includegraphics[scale=0.34, angle=-90]{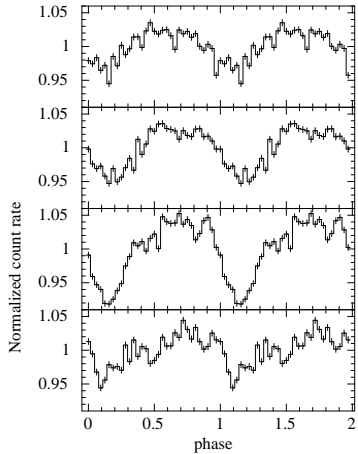}
 \caption{Normalized pulse profiles of March 5 light curve folded at the pulse period shown at 4 different orbital phases
 considering an orbital period 2958 s. }
 \end{figure}
 
The \software{efsearch} test carried out on the light curve of the second night produced several peaks, no single one stands 
out like that from the data of March 5. We have then searched for coherent pulsations in small segments of the
light curve in a wider trial period range and the result is shown in Figure 3 (right-hand panel). Clear pulsations
are detected in at least five segments, but with a pulse period which differ by as much as 0.1 s.
The pulse period monitoring of 4U 1626-67 with the \textit{Fermi} Gamma-Ray Burst Monitor (GBM) shows a period change of the 
order of a few micro-seconds per day in the current spin-up phase, which is much smaller than the period measurement
error from the first night's light curve. The true spin period of the neutron star on the two nights must be quite
similar and we have therefore created pulse profiles from the March 6 light curve with the same period. \\
To study the pulse shape variation as a function of time, the V-band light curves of March 5 and 6 each
were folded on 7.677 s pulse period. A total of 20 and 27 pulse profiles were created for the two nights respectively
which are shown in Figure 6.
Each pulse profile, corresponding to a time segment of $\sim$244 s, was given count rate offset and plotted below the previous profile. 
The segment-wise pulse profiles of March 5 shows no significant drift. However, the March 6 data show a phase
shift in the pulse profile peak as we go along the light curve segments at the rate of 3 pulses per 0.02 days, 
corresponding to a pulse period difference of about 0.1 s as is also seen in pulse period of the second night
(Figure 3, right-hand panel).
We have also examined if the phase drift seen on March 6 night was due to the overall intensity variation in this light curve.
The analysis was repeated after de-trending the light curve with a fourth-order polynomial. The pulse peak drifting effects
were found to be same in the de-trended light curve. \\

\begin{figure*}
\centering
 \includegraphics[scale=0.68, angle=-90]{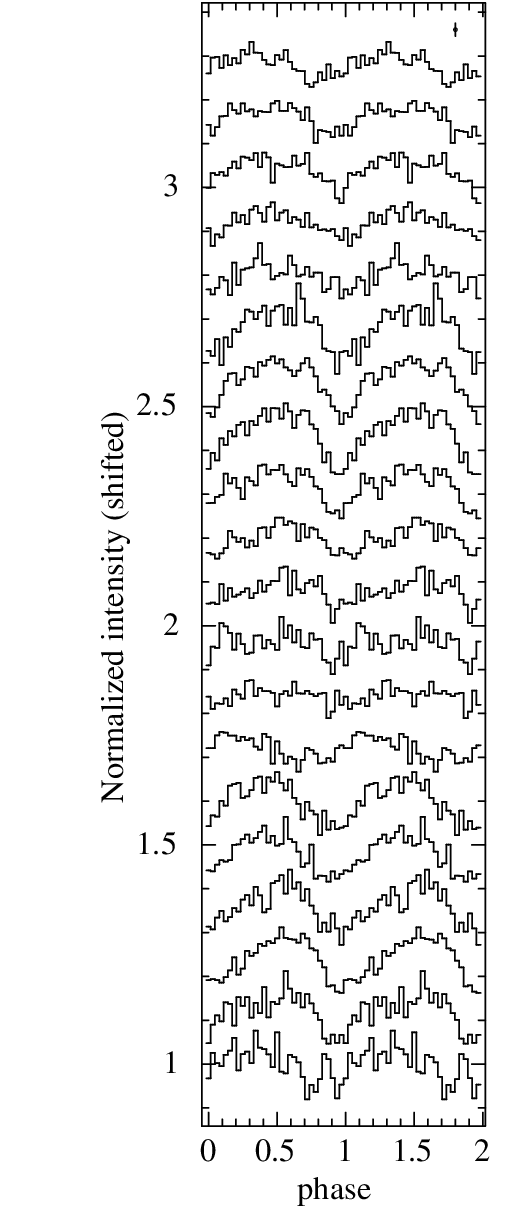}
 \includegraphics[scale=0.68, angle=-90]{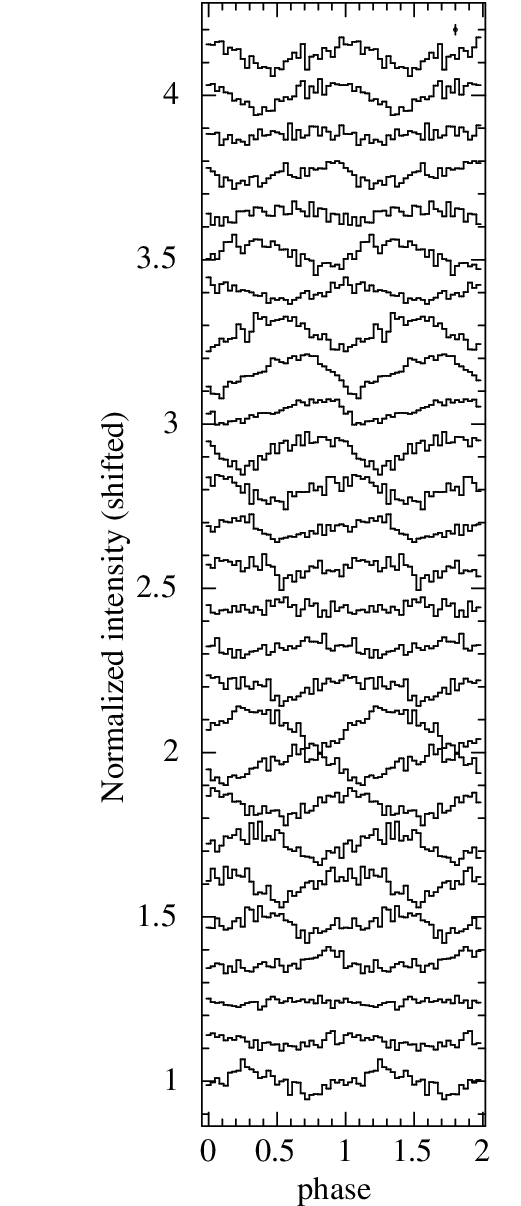}
 \caption{Pulse profile evolution as a function of time for March 5 and 6 nights are shown in the left and right panel,
 respectively. Each pulse profile corresponds to a time segment of $\sim$244 s. 
 To estimate the error, each light curve segment has been folded with a pulse period, away from the main pulsation,
 that showed a very low power in the period search routine. An average standard deviation was then computed for
 the folded profiles from all the segments and is shown in the top-right corner of the figures.}
\end{figure*}

To further probe this drift in the pulse period, we created dynamic power spectrum for segments of 1000 s 
of both the light curves with a moving window of 20 s. The dynamic power spectra are shown in Figure 7 in two frequency
ranges around the fundamental and first harmonic of the pulse frequency. PDS of the 5 night shows persistent features
at the fundamental and the harmonic with lower power at around 2/3rd of the observation. This is consistent with a
smaller pulse fraction in the pulse profiles at that time.
The dynamic power spectrum of March 6 shows many interesting patterns of decreasing pulse frequency with
lower power in the last 20$\%$ of the observation and at around 1/5th of the observation. 
It also shows an upward turn in the frequency near the discontinuity. 
Two frequencies are present simultaneously at about 1/4 th and 3/4 th of the observation. 

 \begin{figure*}
 \centering
\begin{minipage}[c]{\textwidth}
\includegraphics[width=3.5in, angle=0]{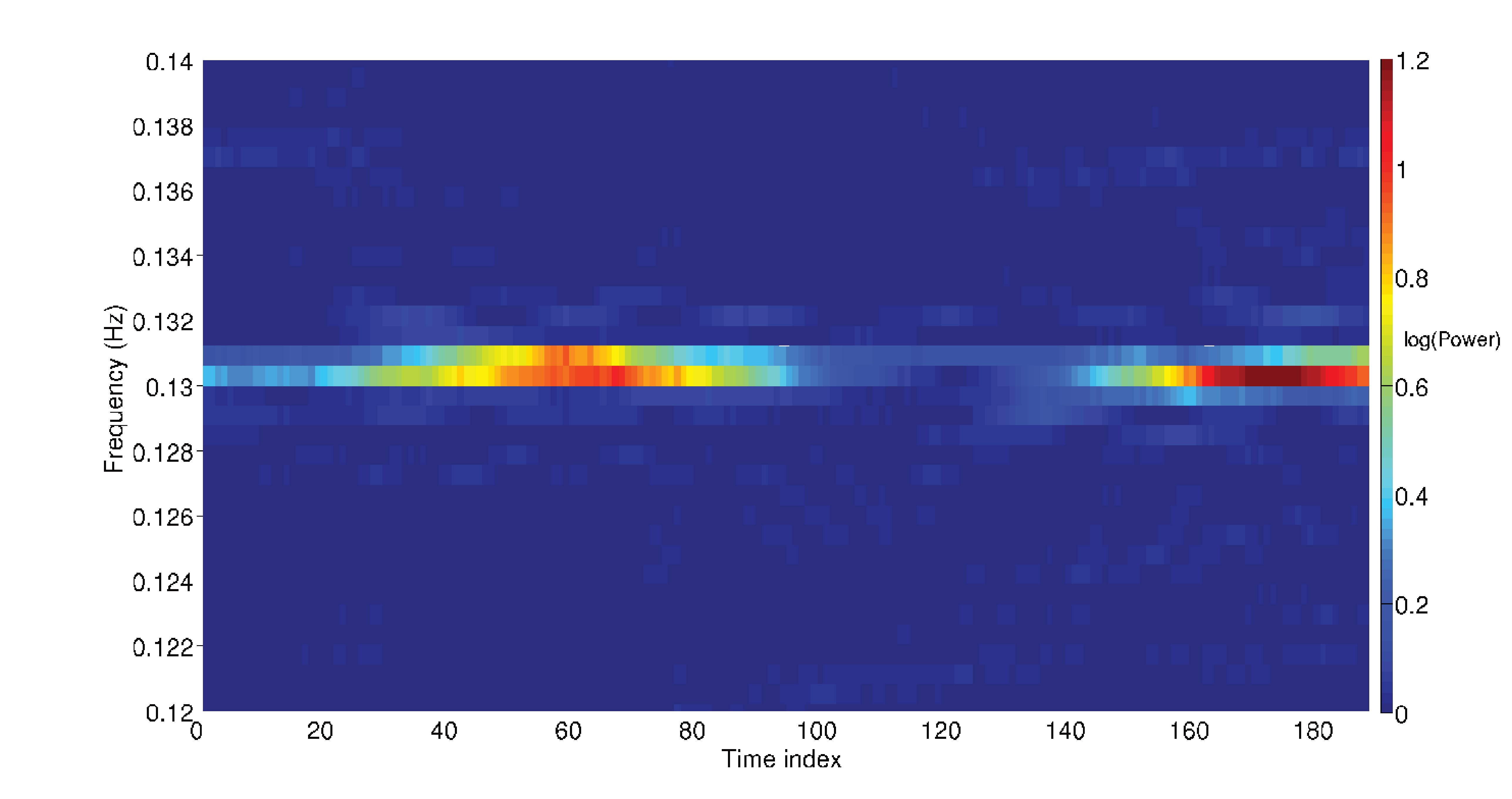}
\includegraphics[width=3.5in, angle=0]{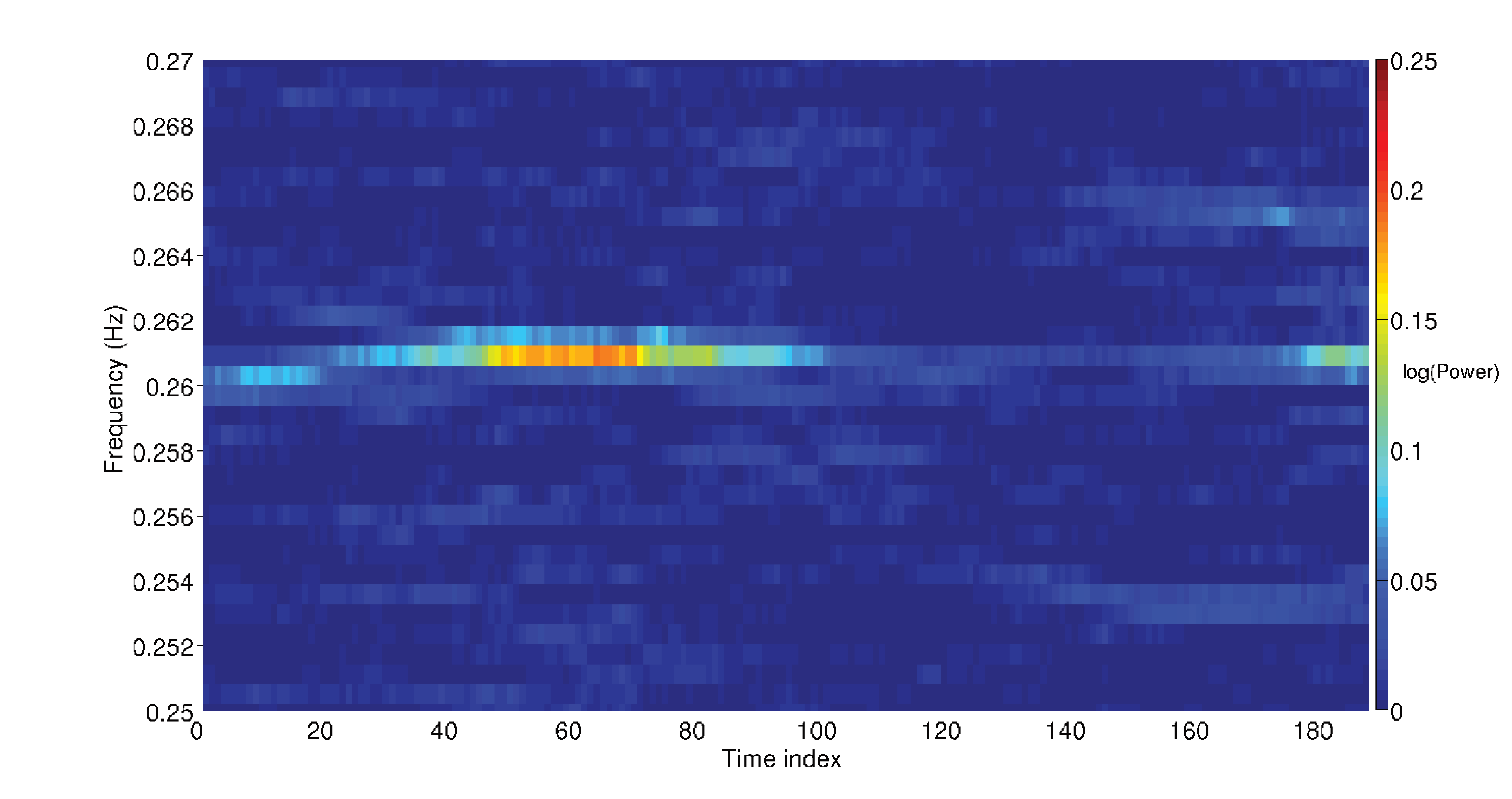}
 \end{minipage}     
\begin{minipage}[c]{\textwidth}
\includegraphics[width=3.5in, angle=0]{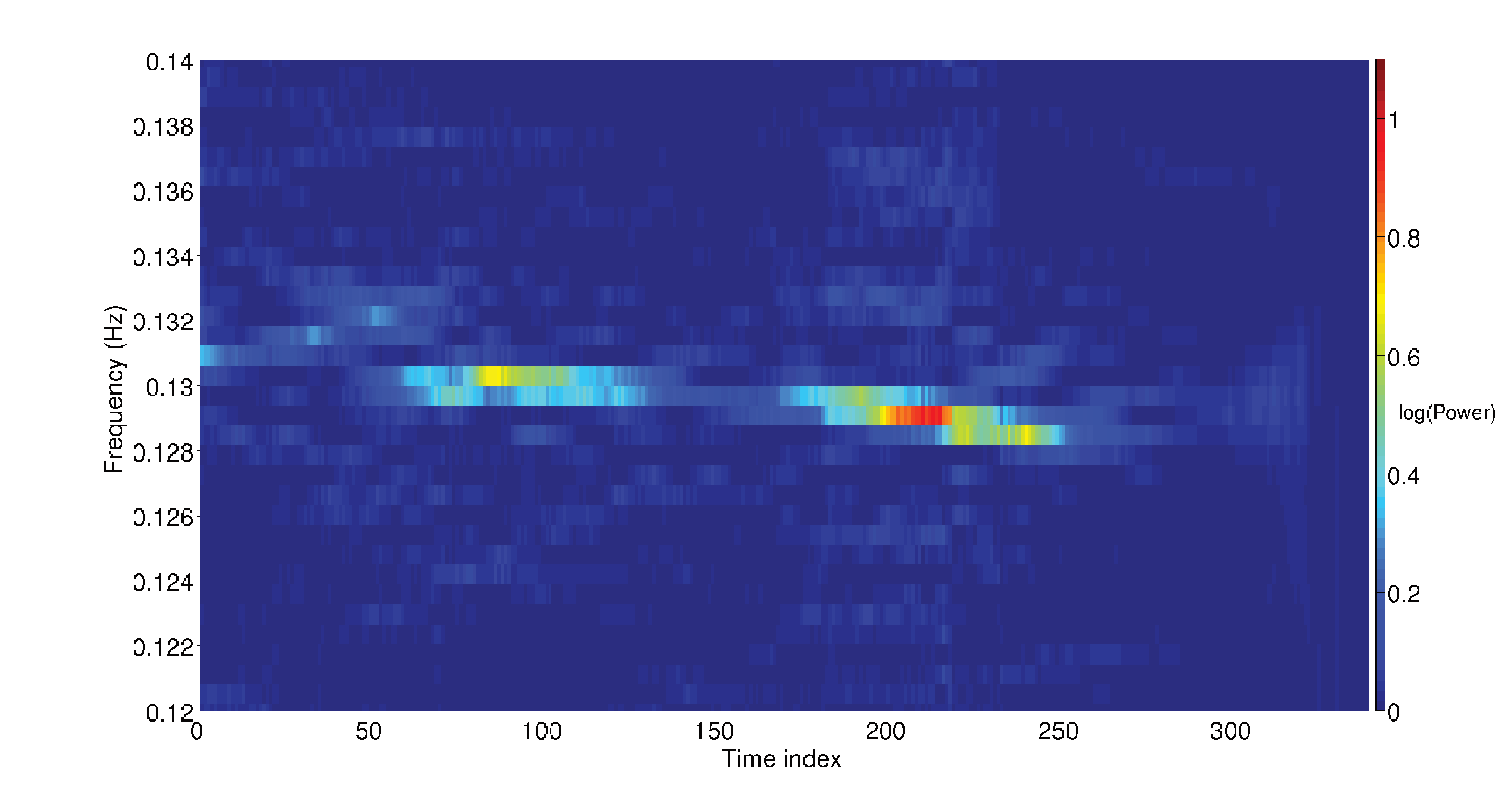}
\includegraphics[width=3.5in, angle=0]{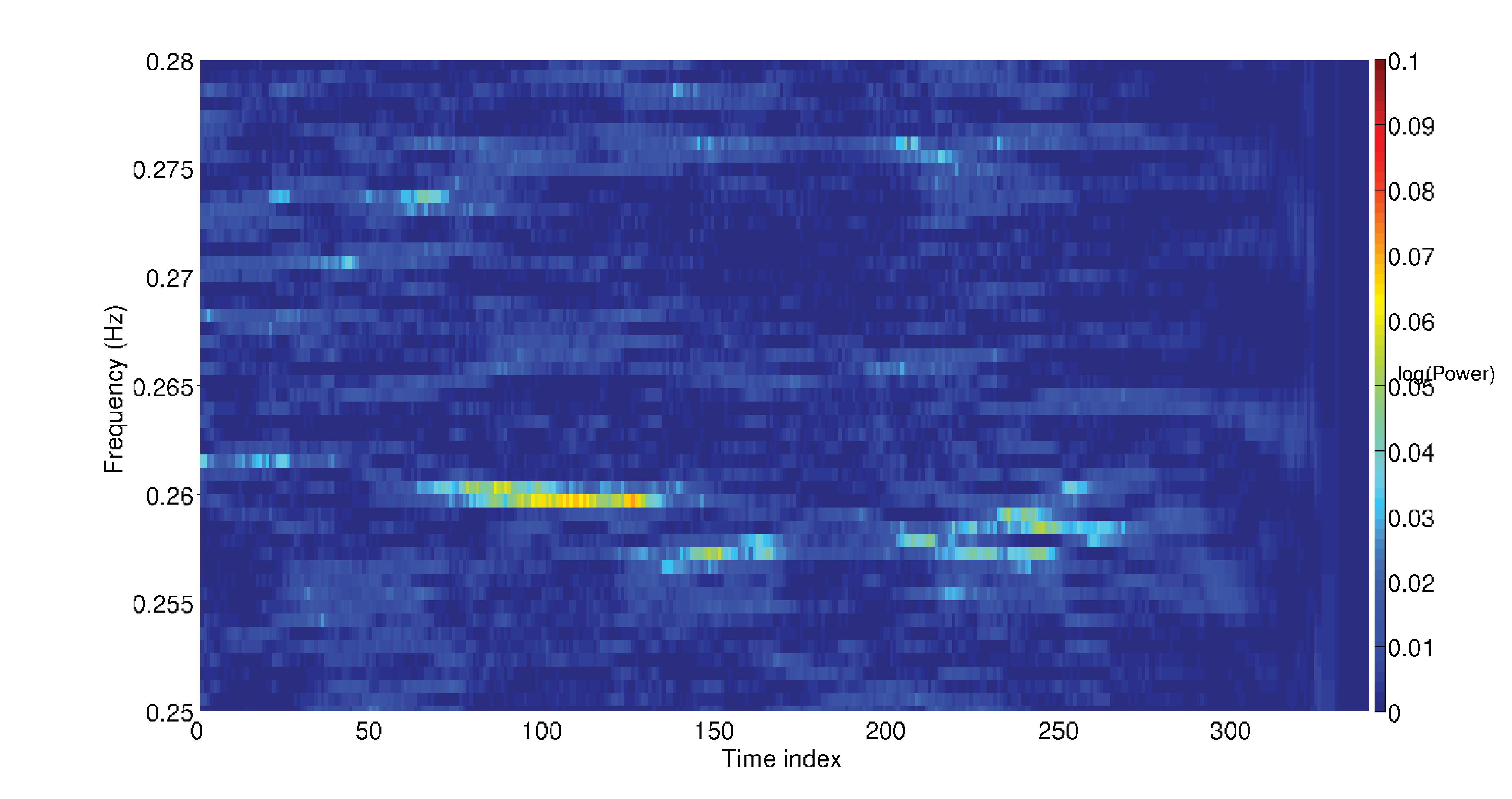}
 \end{minipage}  
\caption{Dynamic power spectra around the fundamental (left) and first harmonic (right)
of the pulse frequency on the nights of March 5 (top) and March 6 (bottom).
Each segment corresponds to 1000 s of data and a moving window of 20 s. 
A clear evolution of the spin period strength and frequency is observed on March 6.}
 \end{figure*}

\subsection{Simulation of the orbital sideband}
The peak of the side-band feature seen from the observation on the night of March 5 has an offset of 0.338 mHz which is
different from values of the same measured earlier \citep{middleditch,deepto1998}. We carried out a simulation to find 
the significance of this difference. Gaussian randomized light curves were simulated at the same time bins as the
observed data using the main band and side-band pulse profiles as shown in Figure 4 (left and right-hand panels) with random phase offset.
Period searches were carried out on 10000 simulated light curves and we found that the short duration of the observation, covering only
about two orbital cycles, is insufficient to constrain the separation between the two components with simulations.
 
\begin{figure}
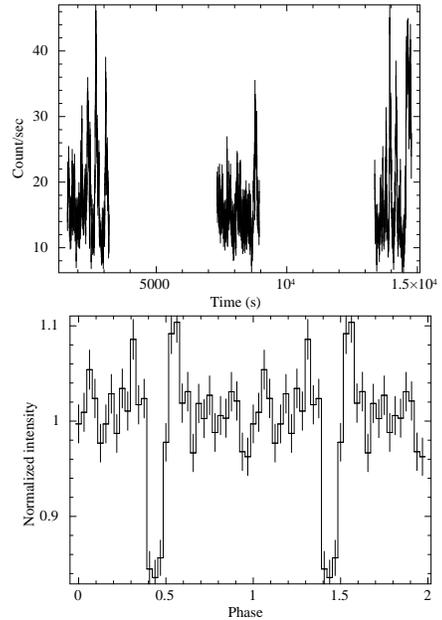

\centering
\includegraphics[scale=0.3,angle=-90]{swift_lc.ps} 
\includegraphics[scale=0.3,angle=-90]{swift_efold.ps} 
\caption{7 s binned \textit{Swift}-XRT light curve (Start time 56721.52 MJD ) (top) 
and the pulse profile folded at the best period 7.6736 s (bottom).}
\end{figure}
 
\subsection{\textit{Swift}-XRT data}
A Target of Opportunity X-ray near-simultaneous observation was carried out on 2014 March 5 (Table 1) using \textit{Swift}-XRT. 
The XRT aboard the \textit{Swift} satellite was operated in the Window Timing mode and given an exposure of 4594 s with a time 
resolution of 0.1 s. The raw data were processed using \software{xrtpipeline} task and using standard screening criteria. 
The source image and light curve were extracted using \software{xselect} in the energy band 0.3 to 10 keV from a circular region
centred on the source co-ordinates. The background subtracted  \textit{Swift} light curve with a bin size of 10 s is shown in Figure 8
(top left panel). The \software{efsearch} task was used to determine the pulse period and a sharp peak was obtained at
7.6736 s (Figure 8 top right panel). The light curve was folded using this pulse period. The characteristic double-horned profile 
is discernible (Figure 8 bottom panel). The \textit{Swift} observation was carried out to know if the properties of the
X-ray source (pulse profile, power density spectrum) were the same as the other X-ray observations in the spin-up phase.
 
\subsection{\textit{RXTE} observation of pulse profiles in flaring state}
The near simultaneous \textit{Swift}-XRT observation which was carried out on March 5 showed no flares and has limited statistics.
So we could not separately investigate intrinsic variations in the X-ray pulse profile using the \textit{Swift}-XRT light curve.
We analysed archival data of the Proportional Counter Array (PCA) on board \textit{RXTE}. The standard 1 light curve in the 
2-60 keV band was extracted for all the observations after the second torque reversal in 2008. The PCA light curve with a bin 
size of 7 s for one such Obs-Id 95313-01-01-08 (Table 1) is shown in Figure 9. The light curve shows intermittent flares.
However, in contrast with the March 6 optical data, period search on the X-ray light curve gives a single sharp peak 
at the pulse period of 7.67 s. The light curve from small segments was folded at this pulse period and
plotted above one another by giving them a finite offset (Figure 10). The X-ray profile is dominated by dips, 
which are locked in phase. They are probably produced by absorption in the accretion stream/column. Flaring portions of the X-ray
light curve seem to change the amplitude of the peaks in the X-ray pulse profiles to some extent, 
but no phase shift is seen in the X-ray pulse profile. We also analysed the PDS from two of the data sets 
from the post second torque-reversal era that had large exposure times and did not have flares in them (Obs Id 95338-05-01-00 and 
95338-05-02-00). The power spectra were fit between the frequency range 2 to 200 mHz with a power law continuum 
and a Lorentzian at the same QPO frequency as obtained from the current SALT dataset. We obtain an upper limit to the rms power of 
the X-ray QPO of 0.9$\%$.

\begin{figure}
\centering
  \includegraphics[scale=0.34,angle=-90]{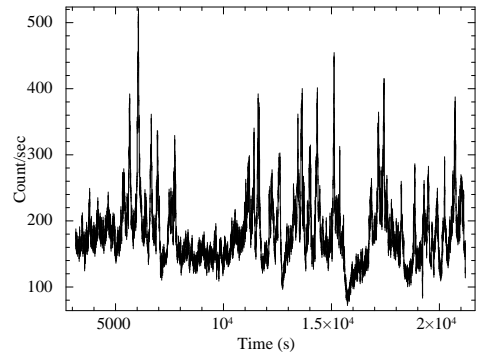}
  \caption{\textit{RXTE}/PCA light curve with a bin size of 7 s corresponding to Obs-Id 95313-01-01-08 showing many flares.} 
\end{figure}
 
 \begin{figure}
 \centering
  \includegraphics[scale=0.6, angle=-90]{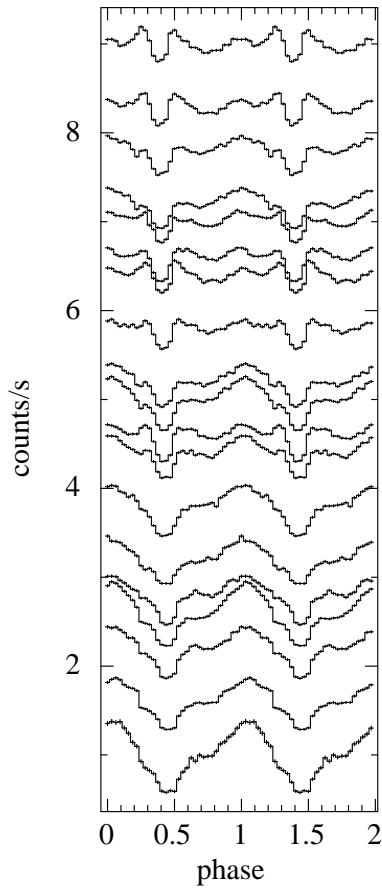} 
   \caption{Time evolution of RXTE pulse profiles folded at the 7.676 s pulse period along the flaring light curve.
   Changes in pulse profile shape is seen but not any phase shift.}
 \end{figure}

\section{Discussion}
4U 1626-67 is an X-ray pulsar that has been found to consistently display timing signatures in the optical that mirror
the X-rays timing features due to reprocessing phenomenon. Our current results indicate many reprocessed X-ray 
signatures like optical pulsations at the spin frequency of the neutron star, the orbital side band due to reprocessing 
of the \textbf{X-ray pulse} from the surface of the companion star, optical flares of a few hundred seconds time-scales,
optical QPOs at 48 mHz and drift in the phase of the reprocessed optical pulses in data from one night.
The accretion disc contributes to most of the  reprocessing that is observed, with the companion star surface also partly
being responsible for reprocessing the main X-ray pulsation. The optical emission from the disc 
arises from a combination of viscous dissipation and X-ray heating \citep{deepto2001}. \\
A 48 mHz QPO was detected simultaneously in the X-ray and optical bands in the spin-down phase right after the
first torque reversal \citep{deepto1998}. Although the first spin-up phase showed a weak QPO at $\sim$40 mHz in X-rays \citep{shinoda}, 
it is absent in the X-ray light curves of the current spin-up phase \citep{beri}.  Interestingly, current SALT results point towards 
the presence of the same 48 mHz QPO in the optical. \\
The presence of the optical 48 mHz QPO at the same X-ray QPO frequency indicates that the X-ray signal
is being reprocessed. However, detection of the QPO feature in the optical light curve with an rms of ~3$\%$,
while the non-simultaneous X-ray light curves with the \textit{RXTE}-PCA in the current spin-up phase having an upper
limit of 1$\%$ in the QPO rms at that frequency indicates various possibilities. There could be an underlying
X-ray QPO which is anisotropic and the optical QPO is produced in a reprocessing region which has different
visibility of the X-ray QPOs than the observer. Or, there is a second possibility that there is an X-ray QPO
in the spin-up phase, but it has variable rms and during the observation on March 5, the X-ray QPO rms was
higher than usual. A third scenario could be that the QPO generation mechanism in the spin-down phase produces
a larger rms in the optical band.\\
QPOs in high magnetic field pulsars are believed to be produced as a result of inhomogeneity at the magnetosphere-inner
disc connection. There are several models that explain the QPOs observed in accretion powered X-ray pulsars.
In the Keplerian frequency model \citep{vanderklis}, the QPO frequency corresponds to the Keplerian frequency at the Alfv{\'e}n
radius of the magnetosphere and QPO frequency is larger than spin frequency of the neutron star which is not the present
case. The beat frequency model \citep{alparshaham} suggests that QPO is at the beat frequency between the Keplerian
frequency at the Alfv{\'e}n radius of the disc-magnetosphere and the neutron star spin frequency. This would predict the
Keplerian frequency to be, $\nu_k$ = $\nu_{spin}$ + $\nu_{QPO}$ $\sim$0.17 mHz. As shown in \citet{shinoda}, this leads to
a consistent measure of the magnetic dipole moment. This model however, fails to explain the asymmetric side-band
amplitudes. The inner disc hypothesis of QPOs (Keplerian frequency model or beat frequency model) can be tested in pulsars
that show both the cyclotron absorption line and QPO \citep{james}. However, in two of the persistent pulsars with known
cyclotron line energy in which the QPOs have been detected over a range of X-ray luminosity, the QPO 
frequency is found to have little luminosity dependence contrary to what would be expected if they had an inner
disc origin (4U 1626-67: \citealt{kaur}; Cen X-3: \citealt{raichurpaul}). \\
A number of other X-ray pulsars have shown flares at intervals comparable to those in the optical light curve of
the second night which produce mHz QPO features in the PDS: Her X-1 \citep{boroson}, LMC X-4 \citep{moonEiken} and 
4U 0115+63 \citep{soongswank}. A magnetically driven warping/precessing model was used to explain mHz 
variability in the X-rays and UV/optical for these sources \citep{akiko}.
A 1 mHz optical QPO, without a counterpart X-ray QPO at that frequency, was discovered in 2001 \citep{deepto2001}
for 4U 1626-67. Since during spin-down the source did not show 1000 s flares in X-rays, the optical features were
attributed to a possible warp in the system. This current data set, in the spin-up phase does show flares and we
see significant power at a 3.32 mHz in the PDS of the March 6 night data. Optical flares which are reprocessed X-ray 
flares might be responsible for such a feature. We have no way of confirming this since we do not have simultaneous X-ray data for the
second night of SALT observation. But another possible scenario could be that the 3.32 mHz frequency,
like an earlier reported 1 mHz signal, might be a geometric modulation effect.\\
A tantalizing suggestion in our data is of the difference, by about 0.06 mHz, in the
sideband frequency from what was observed earlier \citep{middleditch,deepto1998}.
If interpreted as a beat between the spin and the orbital frequency, a change of this magnitude
is not expected over such a short time-scale. Our result would need a longer stretch of timing
data to confirm, but here we speculate on possible reasons that could contribute to such
a change. Optical reprocessing by a structure in the outer accretion disc, e.g. a short-lived
warp orbiting with a period of 2958 s, could give rise to the observed sideband, and small changes
in the warped structure can then lead to frequency drift of this feature.  Warps in the
accretion discs of LMXBs may be driven by instabilities discussed by \citet{Pringle}. While   
the conditions for excitation are met in limited circumstances \citep{OgliDub},
strong irradiation of the disc in the present system may help drive such behavior.\\
The March 6th light curve showed a definite drift of the phase of the optical pulse with time. The reprocessed optical main
pulsation could show such a pulse phase shift if the original X-ray pulse suffered a similar phase shift in the first place.
From analysis of many \textit{RXTE} observations in the current spin-up phase, we have detected significant changes in the X-ray pulse
profile at short time-scales in terms of changes in the relative intensities of the two peaks. However, the X-ray pulse profiles
obtained with the \textit{RXTE}-PCA light curves including flares do not show pulse phase drift as is seen in the optical
light curve of the second night. The SALT observations therefore do not suggest changes in the rotation period of the neutron star or
the pulse period of the X-ray emission. The light curve folded with the period of the neutron star pulse period 
shows the reprocessed optical emission to be shifting in phase, indicating a changing location of the reprocessing region.
However, the time-scale for this change is smaller, about 600 seconds, compared to the sideband observed in the first night which 
is at a five times longer timescale. The SALT observations therefore probably show warps at different radii in the accretion disc for 
two consecutive nights.\\
Longer, and simultaneous X-ray and UV/optical timing study would yield a more comprehensive understanding of the size, extent
and location of the reprocessing sites within the binary and give more insight into the effects of torque reversal on the inner 
disc-magnetosphere geometry.\\

\textit{Acknowledgements}: 
This work has partly made use of data obtained from the High Energy Astrophysics Science Archive Research Center (HEASARC),
provided by the NASA Goddard Space Flight Center. Some of the observations reported in this paper were obtained with the
Southern African Large Telescope (SALT) under programme 2013-2-IUCAA$\_$OTH-001 (PI: D. Bhattacharya).
The authors are grateful to David Buckley and the SALT team for carrying out the optical fast photometric observations and providing
advice on data analysis techniques. The support of the \textit{Swift} team in scheduling a simultaneous target of opportunity observation is
gratefully acknowledged. We also thank Ms. Aru Beri for useful discussions.


\begin{thebibliography}{}

\bibitem[\protect\citeauthoryear{Alpar \& Shaham}{1985}]{alparshaham}
Alpar, M. A., \& Shaham, J. 1985, Nature, 316,239
\bibitem[\protect\citeauthoryear{Beri et al.}{2014}]{beri}
Beri, A., Jain, C., Paul, B., Raichur, H. 2014, MNRAS, 439, 1940B
\bibitem[\protect\citeauthoryear{Bildsten et al.}{1997}]{bildsten}
Bildsten, L., et al. 1997, ApJS, 113, 367
\bibitem[\protect\citeauthoryear{Boroson et al.}{2000}]{boroson}
Boroson, B., O'Brien, K., Kallman T., Still M., Boyd P. T., Quaintrell H., Vrtilek S.D., 2000, ApJ, 545, 399
\bibitem[\protect\citeauthoryear{Buckley et al.}{2006}]{buckley}
Buckley, D. A. H., Swart, G. P., Meiring, Jacobus, G. 2006, in Stepp L., ed., SPIE Vol. 6267,32
\bibitem[\protect\citeauthoryear{Camero-Arranz et al.}{2010}]{cameroarranz}
Camero-Arranz, A., Finger, M. H., Ikhsanov, N. R., Wilson-Hodge, C. A., Beklen, E. 2010, Apj, 708, 1500
\bibitem[\protect\citeauthoryear{Chakrabarty et al.}{1997}]{deepto1997}
Chakrabarty, D., Bildsten, L., Grunsfeld, J. M., Koh, D. T., prince, T. A., Vaughan, B. A., Finger, M. H., Scott, D. M., Wilson, R. B. 1997, ApJ, 474, 414
\bibitem[\protect\citeauthoryear{Chakrabarty}{1998}]{deepto1998}
Chakrabarty, D. 1998, ApJ, 492, 342
\bibitem[\protect\citeauthoryear{Chakrabarty et al.}{2001}]{deepto2001}
Chakrabarty, D., Homer, L., Charles, P. A., O'Donoghue, D. 2001, ApJ, 562, 985,988-990
\bibitem[\protect\citeauthoryear{Chester}{1979}]{chester}
Chester, T. J.  1979, ApJ, 227, 569, 577 
\bibitem[\protect\citeauthoryear{Crampton \& McClure}{1979}]{crampton}
Crampton, D. \& McClure, R. D. 1979,  Astronomical Society of the Pacific, 91, 118
\bibitem[\protect\citeauthoryear{Gandhi et al.}{2010}]{gandhi}
Gandhi, P., Dhillon, V. S., Durant, M., Fabian, A. C., Kubota, A., Makishima, K., Malzac, J., Marsh, T. R., Miller, J. M., Shahbaz, T., Spruit, H. C., Casella, P. 2010, MNRAS, 407, 2166
\bibitem[\protect\citeauthoryear{G\"{u}nther}{2010}]{gunther}
G\"{u}nther, H. M., Lewandowska, N., Hundertmark, M. P. G., Steinle, H., Schmitt, J. H. M. M., Buckley, D., Crawford, S., O'Donoghue, D., 
Vaisanen, P. 2010, A \& A, 518, 54
\bibitem[\protect\citeauthoryear{Hynes et al.}{2006}]{hynes}
Hynes, R. I., Horne, K., O'Brien, K., Haswell, C. A., Robinson, E. L., King, A. R., Charles, P. A., Pearson, K. J. 2006, ApJ, 648, 1156
\bibitem[\protect\citeauthoryear{Ilovaisky et al.}{1978}]{IlovaskyMotchChev}
Ilovasky, S. A., Motch, C., Chevalier, C. 1978, A \& A, 70, L19
\bibitem[\protect\citeauthoryear{Jablonski et al.}{1997}]{jablonski}
Jablonski, F. J., Pereira, M. G., Braga, J., Gneiding, C. D. 1997, ApJL, 482, L171
\bibitem[\protect\citeauthoryear{Jain et al.}{2008}]{jain2008}
Jain, C., Paul, B., Joshi, K., Dutta, A., Raichur, H. 2008, J. Astrophys. Astron., 28, 175
\bibitem[\protect\citeauthoryear{Jain et al.}{2009}]{jain}
Jain, C., Paul, B., Dutta, A. 2009, MNRAS, 403, 920
\bibitem[\protect\citeauthoryear{James}{2013}]{james}
James, M. 2013, PhD thesis, Mahatma Gandhi Univ., Kottayam
\bibitem[\protect\citeauthoryear{Joss et al.}{1978}]{jar}
Joss, P. C., Avni, Y., Rappaport, S. 1978, ApJ, 221, 645
\bibitem[\protect\citeauthoryear{Kaur et al.}{2008}]{kaur}
Kaur, R., Paul, B., Kumar, B., Sagar, R. 2008, ApJ, 676, 1184
\bibitem[\protect\citeauthoryear{Levine et al.}{1988}]{levine}
Levine, A., MA, C. P., McClintock, J., Rappaport, S., van der Klis, M., Verbunt, F. 1988, ApJ, 327,732
\bibitem[\protect\citeauthoryear{McClintock et al.}{1977}]{mcclintock1977}
McClintock, J. E., Canizares, C. R., Bradt, H. V., Doxsey, R., E., Jernigan, J. G. 1977, Nature, 270, 320
\bibitem[\protect\citeauthoryear{McClintock et al.}{1980}]{mcclintock1980}
McClintock, J. E., Canizares, C. R., Li, F. K., Grindlay, J. E. 1980, ApJ, 235,L81,L83-L85
\bibitem[\protect\citeauthoryear{Middleditch et al.} {1981}]{middleditch}
Middleditch, J., Mason, K. O., Nelson, J.E., White, N.E. 1981, ApJ, 224, 1001, 1017
\bibitem[\protect\citeauthoryear{Moon and Eikenberry}{2001}]{moonEiken}
Moon, D., Eikenberry, S. S. 2001, ApJ, 549:L225
\bibitem[\protect\citeauthoryear{Motch et al.}{1982}]{mci}
Motch, C., Ilovaisky, S. A., Chevalier, C. 1982, A \& A, 109,L1
\bibitem[\protect\citeauthoryear{O'Donoghue et al.}{2006}]{donoghue}
O'Donoghue, D., Buckley, D. A. H., Balona, L. A., et al. 2006, MNRAS, 372, 151
\bibitem[\protect\citeauthoryear{Ogilvie and Dubus}{2001}]{OgliDub}
Ogilvie, G. I., Dubus, G. 2001, MNRAS, 320, 495, 496
\bibitem[\protect\citeauthoryear{Orlandini et al.}{1998}]{orlandini}
Orlandini, M. et al. 1998, ApJL, 500, L163
\bibitem[\protect\citeauthoryear{Paul et al.}{2012}]{bpaul}
Paul, B., Archana, M., Saripalli, L. 2012, Bull. Astro. Soc. India, 40, 93
\bibitem[\protect\citeauthoryear{Pringle}{1996}]{Pringle} 
Pringle, J. E. 1996, MNRAS, 281, 357-361
\bibitem[\protect\citeauthoryear{Raichur and Paul}{2008}]{raichurpaul}
Raichur, H. \& Paul, B. 2008, MNRAS, 387, 439
\bibitem[\protect\citeauthoryear{Rappaport et al.}{1977}]{rappaport}
Rappaport, S., Markert, T., Li, F. K., Clark, G. W., Jernigan, J. G., McClintock, J. E. 1977, ApJ, 217,L29
\bibitem[\protect\citeauthoryear{Shinoda et al.}{1990}]{shinoda}
Shinoda, K., Kii, T., Mitsuda, K., Nagase, F., Tanaka, Y., Makishima, K., SHibazaki, N. 1990, PASJ, 42, L27
\bibitem[\protect\citeauthoryear{Shirakawa and Lai}{2002}]{akiko}
Shirakawa, A., Lai, D. 2002, ApJ, 565, 1134
\bibitem[\protect\citeauthoryear{Soong and Swank}{1989}]{soongswank}
Soong, Y., \& Swank, J. H. 1989, in White N. E., Junt J. J., Battrick B., eds, 
Proc. 23rd ESLAP Symp., Two topics in X-ray Astronomy, p. 617
\bibitem[\protect\citeauthoryear{van der Klis et al.}{1987}]{vanderklis}
van der Klis, M., Stella, L., White, N., Jansen, F., Parmar, A. N. 1987, ApJ, 316, 411

\end{thebibliography}
\end{document}